\documentclass[aps,prb,twocolumn]{revtex4}
\usepackage{amssymb}
\usepackage{graphicx}
\usepackage{hyperref}

\begin{document}
\title{Unusual Magnetic, Thermal, and Transport Behaviors of Single Crystal EuRh$_2$As$_2$}
\author{Yogesh Singh, Y. Lee, B. N. Harmon, and D. C. Johnston}
\affiliation{Ames Laboratory and Department of Physics and Astronomy, Iowa State University, Ames, Iowa 50011}
\date{\today}

\begin{abstract}
An \emph{antiferromagnetic} transition is observed in single crystal EuRh$_2$As$_2$ at a high temperature $T_{\rm N} = 47$~K compared to the \emph{ferromagnetic} Weiss temperature $\theta = 12$~K\@.  We show that this is, surprisingly, consistent with mean field theory.  A first-order field-induced magnetic transition is observed at $T < T_{\rm N}$ with an unusual temperature dependence of the transition field.  A dramatic magnetic field-induced reduction of the electronic specific heat coefficient at 1.8--5.0~K by 38\% at 9~T is observed.  In addition, a strong positive magnetoresistance and a large change in the Hall coefficient occur below 25~K\@.  Band structure calculations indicate that the Fermi energy lies on a steep edge of a narrow peak in the density of states.

\end{abstract}

\maketitle

The recent discovery of superconductivity with transition temperatures up to $T_{\rm c} = 38$~K in the layered iron arsenides $A$Fe$_2$As$_2$ ($A$~=~Ba, Sr, Ca, and Eu) when the $A$ atoms are partially replaced by K (Ref.~\onlinecite{Sadovskii2008}) has led to a renewed interest in ThCr$_2$Si$_2$-structure materials.  We have been carrying out a search of similar isostructural compounds such as Ba(Rh,Mn)$_2$As$_2$ (Ref.~\onlinecite{Singh2008}) in an attempt to significantly increase the maximum $T_{\rm c}$ for this class of compounds.  Nature provided a gift of a different sort when we studied the physical properties of another member\cite{Hellmann2007} of this structure class, EuRh$_2$As$_2$, and found a variety of novel behaviors as reported here.

Our primary results are as follows.  First, from our anisotropic magnetic susceptibility $\chi$ versus temperature $T$ data on EuRh$_2$As$_2$ single crystals, the Eu ions are found to have an intermediate valence 2.13(2) unusually close\cite{Matsuda2008} to Eu$^{+2}$, which has a spin-only magnetic moment with $J = S = 7/2$.  Second, an unusually large \emph{antiferromagnetic} ordering temperature $T_{\rm N} = 47$~K compared to the \emph{ferromagnetic} (positive) Weiss temperature $\theta \approx$~12~K is found.  It is widely assumed that the magnitude of $\theta$ in the Curie-Weiss law $\chi=C/(T-\theta)$ is the mean-field transition temperature for either ferromagnetic FM or antiferromagnetic AF ordering of a local moment system, which is the maximum transition temperature that the system can have.  Magnetic fluctuations and frustration effects reduce the magnetic ordering temperature below the mean-field value, so our observation that $T_{\rm N}/|\theta| \approx 4 \gg 1$ is surprising.  The resolution of this conundrum is simple: mean-field theory for a local moment antiferromagnet in fact \emph{allows arbitrarily large values} of the ratio $T_{\rm N}/|\theta|$.\cite{exclude}  This can happen in an antiferromagnet when FM exchange interactions between spins within the same sublattice exist, in addition to the usual AF interactions between spins on opposite sublattices.  

Third, a very unusual and dramatic monotonic magnetic field-induced reduction of the electronic specific heat coefficient $\gamma$ is observed at 1.8--5.0~K by 38\% at a relatively low field of 9~T\@.  We suggest that field-induced stabilization\cite{Matsuda2008} of the +2 valence of Eu is centrally involved.  Finally, a strong positive magnetoresistance develops below 25~K that violates Kohler's rule, where $\rho(T)$ shows a ``nonmetallic'' increase with decreasing $T$ at fixed $H$, together with a large change in the Hall coefficient below 25~K\@.  These apparently coupled electronic behaviors have no obvious origin.  Our band structure calculations indicate that the Fermi energy lies on a steep edge of a sharp peak in the density of states.

Single crystals of EuRh$_2$As$_2$ were grown out of Pb flux.\cite{Hellmann2007}  Single crystal x-ray diffraction measurements confirmed that EuRh$_2$As$_2$ crystallizes in the tetragonal ThCr$_2$Si$_2$ structure with lattice parameters $a = 4.075(4)$~\AA\ and $c = 11.295(2)$~\AA\ at 298~K\@.  The compositions of two crystals were determined using energy dispersive x-ray analysis, yielding the average atomic ratios Eu:Rh:As~$=$~20.8~:~37.9~:~41.3.  The $\chi(T)$ and magnetization $M$ versus applied magnetic field $H$ isotherms were measured with a Quantum Design MPMS SQUID magnetometer.  The $\rho(T)$, $C(T)$ and Hall effect were measured using a Quantum Design PPMS instrument.  

\begin{figure}[t]
\includegraphics[width=2.6in]{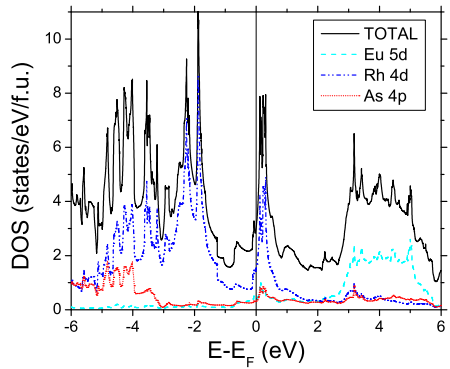}
\caption{ The total density of states DOS for EuRh$_2$As$_2$ versus energy $E$ relative to the Fermi energy $E_{\rm F}$ and the partial DOS versus $E$ from the Eu, Rh, and As atoms.
\label{FigDOS}}
\end{figure}

For the electronic density of states (DOS) calculations, we used the full potential linearized augmented plane wave method with a local density approximation functional.\cite{Perdew1992}  The difference in energy of 0.01~mRy/cell between successive iterations was used as a convergence criterion.  The employed muffin tin radii are 2.5, 2.2 and 2.2 atomic units for Eu, Rh, and As, respectively.  4$f$ electrons of Eu were treated as core electrons.  The structural data were taken from Ref.~\onlinecite{Hellmann2007}.  The total DOS for both spin directions for EuRh$_2$As$_2$ and the partial DOS for Eu 5$d$, Rh 4$d$ and As 4$p$ electrons versus the energy $E$ relative to the Fermi energy $E_{\rm F}$ are shown in Fig.~\ref{FigDOS}.  $E_{\rm F}$ is located just below an extremely sharp peak in the DOS.  The total DOS at $E_{\rm F}$ is $N(E_{\rm F}) = 3.38$~states/eV~f.u. (f.u. means formula unit) for both spin directions with maximum contribution from the Rh 4$d$ orbitals.

\begin{figure}[t]
\includegraphics[width=2.6in]{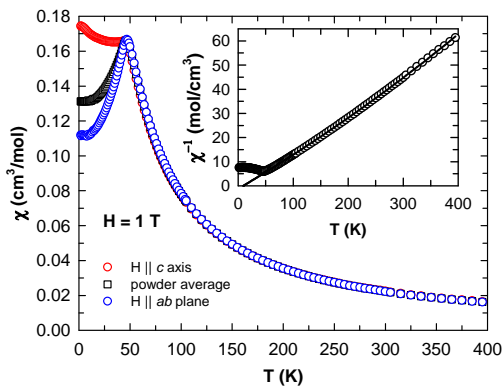}
\caption{$\chi_{\rm ab}$ and $\chi_{\rm c}$ versus temperature $T$ for EuRh$_2$As$_2$.  The powder-averaged $\chi_{\rm powder}$ is also shown.  Inset: fit (solid curve) of the $\chi^{-1}(T)$ data (open circles), see text.
\label{Figsus}}
\end{figure}

The $\chi(T)$ data for a crystal of EuRh$_2$As$_2$ measured with $H$ parallel ($\chi_c$) and perpendicular ($\chi_{ab}$) to the $c$ axis are shown in Fig.~\ref{Figsus}.  The powder-averaged susceptibility $\chi_{\rm powder}~=~(2\chi_{ab}+\chi_{c})/3$ is also shown in Fig.~\ref{Figsus}.  The $\chi_{\rm powder}(T)$ data above 60~K were fitted by the expression $\chi(T) =  f\chi_{\rm Eu^{+3}}(T) + (1-f)C/(T-\theta)$, where the Van Vleck susceptibility $\chi_{\rm Eu^{+3}}(T)$ of Eu$^{+3}$ is given in Ref.~\onlinecite{VanVleck}, $C$ is the Curie constant for Eu$^{+2}$ with $g$-factor $g=2$,\cite{Taylor1975} and $\theta$ is the Weiss temperature for interactions between Eu$^{+2}$ moments.  An excellent fit was obtained with $f = 0.13(2)$ and $\theta = 12(2)$~K (inset).  An average valence of 2.13(2) is therefore obtained for Eu.  This is different from the value $\approx 2.00$ obtained for EuRh$_2$As$_2$ in Ref.~\onlinecite{Michels1996}, possibly due to composition differences of the samples.

The positive value $\theta = 12$~K indicates predominantly \emph{ferromagnetic} exchange interactions between the magnetic Eu$^{2+}$ moments.  Surprisingly, however, in Fig.~\ref{Figsus} we observe a sharp decrease in $\chi_{ab}$ indicating a transition into an \emph{antiferromagnetic} state at a much \emph{higher} N\'eel temperature $T_{\rm N} = 47$~K\@.  The $\chi_{c}$ also shows an abrupt change in slope at $T_{\rm N}$ and becomes weakly temperature dependent at lower $T$.  The large value of $\chi_{ab}(T\to 0)$ indicates that EuRh$_2$As$_2$ is a noncollinear easy plane antiferromagnet with the easy plane being the $ab$ plane.  Magnetic x-ray scattering measurements on our crystals at $H=0$ revealed both commensurate and incommensurate magnetic structures in which the Eu spins are ferromagnetically aligned within the $ab$ plane and where the spins in adjacent planes are, or are nearly, antiparallel.\cite{Nandi2009}
 
A large ratio of $T_{\rm N}/|\theta|$ can occur within mean-field theory for a two-sublattice collinear antiferromagnet with equal numbers of spins on the two sublattices, each with Curie constant $C/2$, as follows.  A spin in each sublattice is assumed to interact with the same number of spins both within its own sublattice and with the other sublattice with mean-field coupling constants $\lambda_1$ and $\lambda_2$, respectively.  Applying the usual mean-field treatment one obtains the Weiss temperature $\theta = C(\lambda_1 + \lambda_2)/2$ and magnetic ordering temperature $T_{\rm N} = C(\lambda_1 - \lambda_2)/2$.  Thus
\begin{equation}
\frac{T_{\rm N}}{\theta} = \frac{\lambda_1 - \lambda_2}{\lambda_1 + \lambda_2} = \frac{{\cal J}_1 - {\cal J}_2}{{\cal J}_1 + {\cal J}_2},
\label{EqTNTheta}
\end{equation}
where ${\cal J}_1$ and ${\cal J}_2$ are the nearest-neighbor exchange coupling constants for two spins in the same and different sublattices, respectively.  If $\lambda_1,\ {\cal J}_1 > 0$ (FM) and $\lambda_2,\ {\cal J}_2 < 0$ (AF), one can obtain arbitrarily large values of the ratio $T_{\rm N}/|\theta|$.  For our case with $T_{\rm N}/\theta \approx 4$, Eq.~(\ref{EqTNTheta}) yields $\lambda_1/\lambda_2 = {\cal J}_1/{\cal J}_2 \approx -5/3$.

\begin{figure}[t]
\includegraphics[width=2.6in]{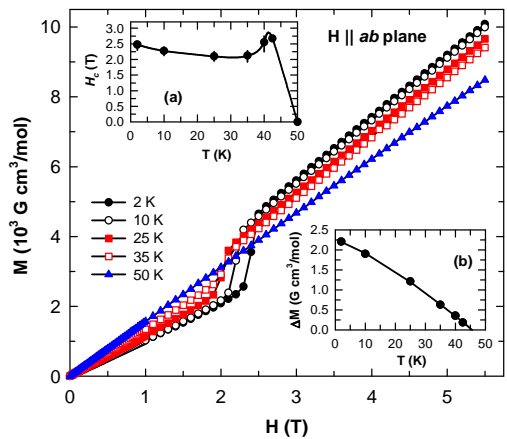}
\caption{$M(H)$ at various $T$ with $H$ applied parallel to the $ab$ plane.  Inset(a): metamagnetic field $H_{\rm c}$ versus $T$.  The vertical bars on the data points are the widths of the metamagnetic transition.  The solid curve is a guide to the eye.  Inset(b): change in magnetization $\Delta M$ at the transition versus $T$. 
\label{FigMH1}}
\end{figure}

$M(H)$ isotherms at various $T$ with $H$ applied along the $ab$ plane are shown in Fig.~\ref{FigMH1}.  The $M(H)$ data for $H \parallel c$ (not shown) are proportional at all temperatures from 2 to 300~K\@.  The $M(H)$ data for $H$ applied along the $ab$ plane are also proportional for temperatures $T > T_{\rm N} = 47$~K as seen in Fig.~\ref{FigMH1}.  However, for $T < T_{\rm N}$ the $M(H)$ is initially proportional but then shows a first-order step-like increase in $M$ at a metamagnetic critical field $H_{\rm c}$ which exhibits hysteresis (not shown) upon increasing and decreasing $H$.  Above $H_{\rm c}$, $M$ again is proportional to $H$ but with a larger slope.  The value of $M$ at $T = 2$~K and $H = 5.5$~T is only 1.81~$\mu_{\rm B}$/f.u., which is much smaller than the expected Eu$^{2+}$ saturation moment $7.0~\mu_{\rm B}$/Eu.  Our data thus indicate that a first-order transition between two antiferromagnetically ordered states occurs at $H_{\rm c}$.  Figure~\ref{FigMH1} inset(a) shows that $H_{\rm c}$ decreases initially with increasing $T$ between 2~K and 25~K, as expected, but then increases strongly on further approaching $T_{\rm N}$.  At $T$~=~50~K~$> T_{\rm N}$ we did not observe any metamagnetic transition.  The increase in magnetization $\Delta M$ across the metamagnetic transition versus $T$ is shown in Fig.~\ref{FigMH1} inset(b).  In contrast to $H_{\rm c}$, $\Delta M$ shows a monotonic decrease with $T$ and vanishes near $T_{\rm N}$ as expected.  

\begin{figure}[t]
\includegraphics[width=2.6in]{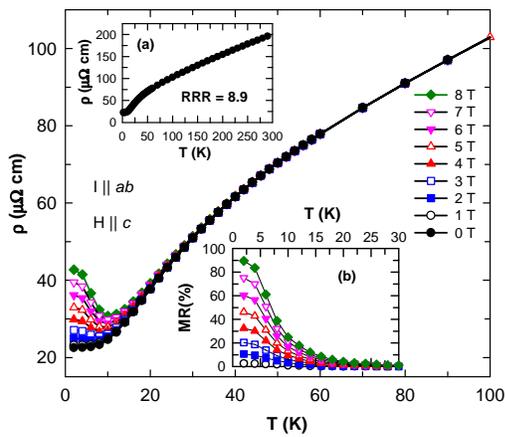}
\caption{Resistivity $\rho$ in the $ab$ plane versus temperature $T$ measured in various $H \parallel c$.  Inset (a): $\rho(T)$ for $H = 0$.  Inset (b): Magnetoresistance MR below $T = 30$~K. 
\label{Figres}}
\end{figure}

The $\rho(T)$ data for current in the $ab$ plane for $H = 0$ and for temperatures from 2~K to 300~K are shown in Fig.~\ref{Figres} inset(a).  These data indicate metallic behavior with a residual resistivity ratio RRR~=~$\rho(300~{\rm K})$/$\rho(2~{\rm K})$~=~8.9.  There is no sudden reduction in $\rho(T)$ below $T_{\rm N}$~=~47~K as might be expected below a magnetic ordering transition due to a reduction of spin-disorder scattering.  This is particularly surprising in view of the sharp transitions at $T_{\rm N}$ seen in $\chi(T)$ and $C(T)$ in Fig.~\ref{Figsus} and in Fig.~\ref{FigHC} below, respectively.

\begin{figure}[t]
\includegraphics[width=2.6in]{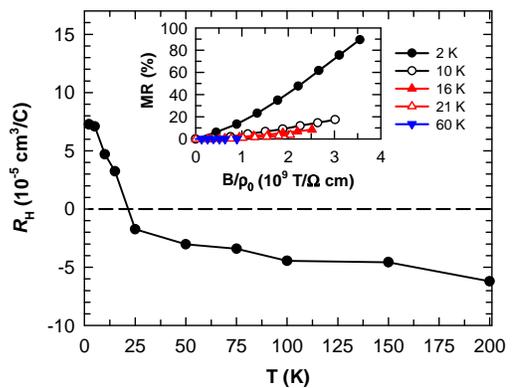}
\caption{Hall coefficient $R_{\rm H}$ vs.\ $T$ for EuRh$_2$As$_2$.  Inset: Magnetoresistance MR versus $H/\rho(H=0)$ at various $T$. 
\label{FigHall}}
\end{figure}

The field-dependent $\rho(T,H)$ data are shown in Fig.~\ref{Figres} between 2~K and 100~K\@.  A strong increase in $\rho$ occurs with increasing $H$ beginning below 25~K\@.  The magnetoresistance percentage values MR($H,T$)~$\equiv$~$100[\rho(H,T)-\rho(0,T)]/\rho(0,T)$ versus $T$ at various $H$ are shown in Fig.~\ref{Figres} inset(b).  A large MR is seen at low $T$ with increasing $H$: the MR reaches 90\% at $T = 2$~K and $H=8$~T\@.  From the single-band relation $\omega_{\rm c}\tau = |R_{\rm H}|H/\rho$, where $\omega_{\rm c}$ is the cyclotron frequency, $\tau$ is the mean-free scattering time of the current carriers and $R_{\rm H}$ is the Hall coefficient, and using our experimental $R_{\rm H}$ (below) and $\rho$ data at 2~K, one finds that our MR data are in the low-field regime $\omega_{\rm c}\tau \sim 0.003 \ll 1$ at 8~T\@.  In this regime one normally expects\cite{Pippard1989} MR~$\sim H^2$ instead of the different behavior we observe in Fig.~\ref{FigHall} inset.  A positive MR can occur due to increased spin-disorder scattering upon suppression of an antiferromagnetic ordering by a magnetic field.\cite{Ren2008}  However, this explanation is untenable here because as shown in Fig.~\ref{FigHC} below, the $T_{\rm N}$ of EuRh$_2$As$_2$ is suppressed to only $\sim 40$~K in $H = 8$~T\@.  Furthermore, one expects a zero MR with $H\parallel c$ ($H\perp$ ordered moment direction) due to AF fluctuations at $T \ll T_{\rm N}$.\cite{Yamada1973} According to semiclassical transport theory, the MR follows Kohler's rule ${\rm MR} = F[{H/ \rho(0)}]$, where $F(x)$ is a universal function for a given material, if there is a single species of charge carrier and the scattering time is the same at all points on the Fermi surface.\cite{Pippard1989}  As shown in Fig.~\ref{FigHall} inset, the MR in EuRh$_2$As$_2$ severely violates Kohler's rule.

The Hall coefficient $R_{\rm H}$ was found to be independent of $H$ up to 8~T and is plotted versus $T$ at $H = 8$~T in Fig.~\ref{FigHall}.  $R_{\rm H}$ is negative and increases slowly with decreasing $T$ from 200~K to 25~K, but then increases rapidly below 25~K, the temperature below which the MR also begins to strongly increase.  An unusual $T$ dependence of $R_{\rm H}$ is sometimes seen across a magnetic transition.\cite{hurd1972}  However, the strong increase in $R_{\rm H}$ for EuRh$_2$As$_2$ occurs below 25~K which is well below $T_{\rm N}(H)$ as shown next.

The $C(T)$ of a single crystal of EuRh$_2$As$_2$ measured between 1.8~K and 70~K in various $H \parallel c$ is shown in Fig.~\ref{FigHC}.  For $H=0$, a second-order anomaly with an onset at 48.3~K and a peak at 44.3~K is observed, from which we estimate $T_{\rm N} \approx 46$~K in agreement with the $T_{\rm N}$ found from our $\chi(T)$ data above.  The $C(T)$ data for a single crystal of BaRh$_2$As$_2$,\cite{Singh2008} also shown in Fig.~\ref{FigHC}, were used to estimate the lattice heat capacity of EuRh$_2$As$_2$. Figure~\ref{FigHC} inset(a) shows $\Delta C(T)$ versus $T$ between 2~K and 100~K, obtained by subtracting the heat capacity of BaRh$_2$As$_2$, adjusted for the molar mass difference with EuRh$_2$As$_2$, from that of EuRh$_2$As$_2$.  $\Delta C(T)$ is consistent with a mean-field transition at $T_{\rm N}$ as follows.  In mean-field theory, the magnitude of the heat capacity jump at $T_{\rm N}$ is given by $\Delta C(T_{\rm N})$~=~${5\over 2}R{(2S+1)^2 -1 \over (2S+1)^2+1}$~=~16.2~J/mol~K$^2$ for $S = 7/2$,\cite{Smart1966} where $R$ is the gas constant. This value is close to that observed in Fig.~\ref{FigHC} inset(a).  Furthermore, the entropy difference $\Delta S(T)$ versus $T$ obtained by integrating the $\Delta C(H = 0, T)/T$ versus $T$, as shown in Fig.~\ref{FigHC} inset(a),  reaches the value $R\ln8$ expected for Eu$^{2+}$ moments ($J=S=7/2$) just above $T_{\rm N}$ after which it becomes nearly $T$ independent.  From the $C(T,H)$ data, one sees that $T_{\rm N}$ decreases by only $\sim 5$~K at 8~T\@.  Thus we infer that the strong positive MR below $\sim 25$~K in Fig.~\ref{Figres} does not result from suppression of $T_{\rm N}$ to these low temperatures.

\begin{figure}[t]
\includegraphics[width=2.6in]{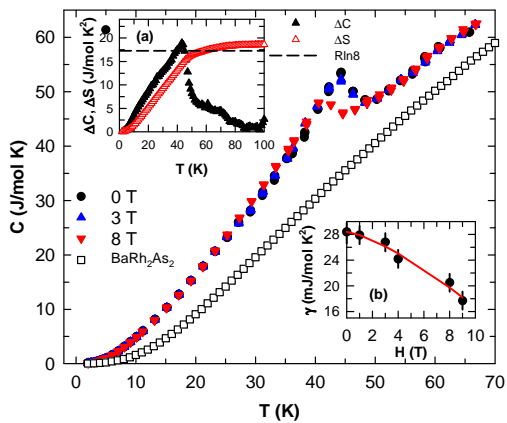}
\caption{Heat capacity $C$ vs.\ $T$ of single crystal EuRh$_2$As$_2$ at various $H\parallel c$, and for single crystal BaRh$_2$As$_2$ in $H = 0$.  Inset~(a): $\Delta C(T)$ and $\Delta S(T)$ vs.\ $T$.  The dashed horizontal line is the value $\Delta S = R\ln 8$ expected for disordered Eu$^{2+}$ ($J = S = 7/2$) spins.  Inset~(b): $\gamma$ versus $H$.
\label{FigHC}}
\end{figure}
\noindent

At 1.8--5.0~K, the heat capacity of EuRh$_2$As$_2$ obeys $C(T,H) = \gamma(H) T + \beta T^3$, where \mbox{$\beta \approx$~7.1(1)~mJ/mol K$^4$} is independent of $H$ and the electronic specific heat coefficient $\gamma(H)$ is plotted in Fig.~\ref{FigHC} inset(b).  Between $H$~=~0~T and 9~T, $\gamma$ decreases monotonically from 28.4(9)~mJ/mol~K$^2$ to 17.7(7)~mJ/mol~K$^2$, a remarkable reduction of 38\%.  This reduction in $\gamma$ might be explained by a field-induced carrier localization; however, the field-independence of $R_{\rm H}$ (above) argues against such an interpretation.  From $N(E_{\rm F}) = 3.38$~states/eV~f.u.\ obtained above from our band structure calculations for a valence Eu$^{+2}$, we obtain $\gamma = 7.96$~mJ/mol~K$^2$ assuming zero electron-phonon coupling.  This value is about 3.5 times smaller than the observed zero-field value.  This discrepancy suggests that the high observed $\gamma(H=0)$ is due to the intermediate valence 2.13(2) of Eu inferred from $\chi(T)$ above $T_{\rm N}$,\cite{varma1976} and that the field-induced reduction in $\gamma$ towards the band structure value arises from field-induced stabilization\cite{Matsuda2008} of the Eu valence towards Eu$^{+2}$ and concomitant reduction in the spin fluctuation\cite{varma1976} contribution.

In summary, our magnetic, transport, and thermal measurements on single crystals of EuRh$_2$As$_2$ revealed an array of interesting and unusual behaviors.  From $\chi(T)$ measurements at temperatures $T > T_{\rm N}$ , the Eu ions are found to have an intermediate valence 2.13(2) unusually close\cite{Matsuda2008} to Eu$^{+2}$.  The large ratio $T_{\rm N}/\theta \approx 4$ is very unusual.  A simple two-sublattice mean-field model where each sublattice interacts with itself in addition to the other explains how $T_{\rm N}/|\theta| > 1$ can come about.  Other relevant examples of antiferromagnets where $T_{\rm N}/|\theta| > 1$ have been reported,\cite{exclude, Levin2004, Budko1999, ibarra1997} although the authors did not take specific note of this ratio. For LaMnO$_3$, using Eq.~(\ref{EqTNTheta}) and the ${\cal J}_{1,2}$ values in Ref.~\onlinecite{McQueeney2008}, one obtains the mean-field ratio $T_{\rm N}/\theta = 3.8$, slightly larger (as expected) than the observed value of 3.0 obtained from $\theta = 46$~K and $T_{\rm N} = 140$~K.\cite{ibarra1997}  In retrospect, it is surprising that antiferromagnets with $T_{\rm N}/|\theta| > 1$ are not more commonly observed.  The temperature variation of the metamagnetic field $H_{\rm c}$ as $T_{\rm N}$ is approached is anomalous.  The strong decrease in the electronic heat capacity coefficient $\gamma$ with $H$ at relatively low fields up to 9~T is very unusual.\cite{EuIrSn}  In most metals, $\gamma$ is independent of $H$ in such fields because the magnetic field energy of a conduction carrier is far smaller than the Fermi energy.  We suggest that the observed $\gamma(H)$ results from a field-induced stabilization\cite{Matsuda2008} of the Eu valence towards Eu$^{+2}$ at low $T$.  This hypothesis can be checked using, \emph{e.g.}, x-ray absorption spectroscopy  (XAS).\cite{Matsuda2008}  A strong positive magnetoresistance and a strong increase in $R_{\rm H}$ develop below 25~K suggesting a possible temperature-induced redistribution of carriers between electron- and hole-like Fermi surfaces, which can be tested using angular-resolved photoemission spectroscopy (ARPES).

\vspace{-0.2in}

\begin{acknowledgments}

We thank S. Nandi, A. Kreyssig, A. I. Goldman and J.~Schmalian for helpful discussions and A. Ellern for structure analysis.  Work at Ames Laboratory was supported by the Department of Energy-Basic Energy Sciences under Contract No.~DE-AC02-07CH11358.  
\end{acknowledgments}


\begin{references}
\bibitem{exclude} We exclude cases where $\theta$ arises from single-ion effects.
\end{references}
\end{document}